\documentclass[fleqn,usenatbib]{mnras}

\usepackage{newtxtext,newtxmath}

\usepackage[T1]{fontenc}
\usepackage{ae,aecompl}

\usepackage{graphicx}	
\usepackage{amsmath}	
\usepackage{gensymb}
\usepackage{scalerel}
\usepackage{tikz}
\usepackage{mathtools}

\defcitealias{Borchert22b}{B22}

\usetikzlibrary{svg.path}

\definecolor{orcidlogocol}{HTML}{A6CE39}
\tikzset{
  orcidlogo/.pic={
    \fill[orcidlogocol] svg{M256,128c0,70.7-57.3,128-128,128C57.3,256,0,198.7,0,128C0,57.3,57.3,0,128,0C198.7,0,256,57.3,256,128z};
    \fill[white] svg{M86.3,186.2H70.9V79.1h15.4v48.4V186.2z}
                 svg{M108.9,79.1h41.6c39.6,0,57,28.3,57,53.6c0,27.5-21.5,53.6-56.8,53.6h-41.8V79.1z M124.3,172.4h24.5c34.9,0,42.9-26.5,42.9-39.7c0-21.5-13.7-39.7-43.7-39.7h-23.7V172.4z}
                 svg{M88.7,56.8c0,5.5-4.5,10.1-10.1,10.1c-5.6,0-10.1-4.6-10.1-10.1c0-5.6,4.5-10.1,10.1-10.1C84.2,46.7,88.7,51.3,88.7,56.8z};
  }
}

\newcommand\orcidicon[1]{\href{https://orcid.org/#1}{\mbox{\scalerel*{
\begin{tikzpicture}[yscale=-1,transform shape]
\pic{orcidlogo};
\end{tikzpicture}
}{|}}}}

\title[FU Orionis-type outbursts from colliding discs]{Sustained FU Orionis-type outbursts from colliding discs in stellar flybys}
\author[Borchert et al.]{
Elisabeth M. A. Borchert\orcidicon{0000-0002-6994-8874},$^{1}$\thanks{E-mail: elisabeth.borchert@monash.edu}
Daniel J. Price\orcidicon{0000-0002-4716-4235},$^{1}$
Christophe Pinte\orcidicon{0000-0001-5907-5179},$^{1,2}$
Nicol\'as Cuello\orcidicon{0000-0003-3713-8073}$^{2}$
\\
$^{1}$School of Physics and Astronomy, Monash University, Vic 3800, Australia\\
$^{2}$Univ. Grenoble Alpes, CNRS, IPAG / UMR 5274, F-38000 Grenoble\\
}

\date{Accepted XXX. Received YYY; in original form ZZZ}

\pubyear{2022}

\begin{document}
\label{firstpage}
\pagerange{\pageref{firstpage}--\pageref{lastpage}}
\maketitle

\begin{abstract}
    We perform 3D hydrodynamics simulations of disc-disc stellar flybys with on-the-fly Monte Carlo radiative transfer. We show that pre-existing circumstellar discs around both stars result in fast rising ($\sim$yrs) outbursts lasting 2--5 times longer than for a star-disc flyby. The perturber always goes into outburst ($\dot{M}>10^{-5}~{\rm M_{\odot}~ yr^{-1}}$). Whereas we find that the primary goes into a decades long outburst only when the flyby is retrograde to the circumprimary disc rotation. High accretion rates during the outburst are triggered by angular momentum cancellation in misaligned material generated by the encounter. A large fraction of accreted material is alien.
\end{abstract}

\begin{keywords}
hydrodynamics --- methods:numerical --- protoplanetary discs --- stars:protostars --- stars: variables: T Tauri, Herbig Ae/Be
\end{keywords}

\section{Introduction}

Young stars undergo unexpected mass accretion rate bursts, most commonly known as FU Orionis outbursts \citep{Herbig66a, Herbig77a, Hartmann96a}. Over the decades since the first detection in FU Ori in 1936, a number of theories have been presented to explain these outbursts. \cite{Lin85a}, \cite{Clarke90a} and \cite{Bell95a} argued for disc thermal instabilities as a more likely scenario, as binary interactions \citep{Bonnell92a, Reipurth04a} were thought to be less likely. This was due to the inability of binary interactions to maintain outbursts for more than a hundred years \citep{Clarke90a}, the assumption that a sudden outburst is only possible with encounters at $<1$~au \citep{Hartmann96a} and a one in three detection of companions \citep{Green16a}. Detection of companions is difficult as it is hard to observe low mass stars directly or indirectly and the companion could have also already left. \cite{Bate18a} showed that almost all young stars undergo stellar interactions in their first Myr.

\cite{Wang04a} first observed that FU Ori itself is actually a binary system instead of a single star. Further observations of FU Ori found evidence for a past encounter as well as the existence of discs around both stars \citep{Reipurth04a, Beck12a, Liu16a, Takami18a, Perez20a}. Z CMa, which also undergoes FU Orionis type outbursts \citep{Hartmann89a,Bonnefoy17a} has been found to have indications of a likely flyby \citep{Dong22a}.

\cite{Vorobyov21a} presented theoretical simulations which show that a disc penetrating stellar flyby leads to accretion bursts of a similar amplitude to those seen in FU Ori. In \citet[][hereafter \citetalias{Borchert22b}]{Borchert22b}, we confirmed these findings using 3D smoothed particle hydrodynamics (SPH) simulations with live Monte Carlo radiative transfer. We further showed that close (10--40 au) disc-penetrating stellar flybys can lead to the short rise times (1--10~years) observed in several FU Orionis objects \citep{Hartmann96a}. Importantly, both studies found that it is not the primary but the secondary star that goes into outburst, as observed in FU Ori \citep{Beck12a,Perez20a} and previously discerned by \cite{Forgan10a}.

The outstanding mystery still to be explained is the long duration of the outburst at a constant level, 86 years and continuing, in FU Ori. 

Our primary hypothesis in this paper is that flyby encounters with two discs will lead to longer lasting outbursts than in a star-disc flyby where only a circumprimary disc exists beforehand. Several authors have previously simulated disc-disc encounters using either SPH \citep{Watkins98a, Watkins98b} or N-body simulations \citep{Pfalzner05a}, showing the re-distribution of mass, formation of shock fronts between the stars, gravitational instability leading to new companions and the capturing of stars. \cite{Picogna14a} also performed SPH simulations of interacting discs, though their focus was on resulting effects on planet orbits and accretion from the disc onto the planet. 

Our second hypothesis is that the pre-existence of a circumsecondary disc would lead to different observational consequences. For example, \cite{Winter18a} discuss the evidence of such a past encounter for HV and DO Tau. Furthermore, a past disc-disc encounter could possibly explain why both stars in FU Ori have compact mm emission \citep{Perez20a}.

To test our hypotheses we perform a series of disc-disc stellar flyby numerical experiments building on the previous scenarios presented in \citetalias{Borchert22b}. We consider the effects of different disc orientations prior to their encounter in comparison with a star-disc flyby. The paper is organised as follows: Section~\ref{sec:methods} explains the methods behind out simulations, Section~\ref{sec:results} presents the results with Section~\ref{sec:discussion} discussing them. Our conclusions are listed in Section~\ref{sec:conclusion}. Accompanying material is presented in the Appendix.

\vspace{-0.25cm}
\section{Methods}\label{sec:methods}
Following the initial flyby setup presented in \cite{Cuello19a}, we performed 3D hydrodynamical simulations with the SPH code \textsc{phantom} \citep{Price18b} with on-the-fly radiative transfer using the Monte Carlo radiative transfer code \textsc{mcfost} \citep{Pinte06a, Pinte09a}. The simulations presented here build on the stellar flyby simulations presented in \citetalias{Borchert22b}. We performed disc-disc flyby encounters, meaning that both stars were assumed to host a pre-existing circumstellar disc at $t=0$. 

The temperatures of the discs were updated with \textsc{mcfost} at set intervals, with the aim of computing disc temperature profiles self-consistently. The intervals were set to 1/100 of the time for the flyby to reach the same separation post periastron that it started at initially (using Barker's equation; \citealt{Cuello19a}), leading to 50 temperature updates on the approach and after periastron with a time resolution of $\sim3.51$~yrs. \textsc{mcfost} takes the gas density structure from \textsc{phantom} and using Voronoi tesselation, each \textsc{phantom} particle is mapped to a corresponding Voronoi cell for the \textsc{mcfost} calculations. With no dust present in the \textsc{phantom} simulations, we assumed (for the purpose of radiative transfer) that the dust distribution followed the gas with a dust-to-gas ratio of 0.01 and a power-law grain size distribution \mbox{${\rm d}n/{\rm d}s \propto s^{-3.5}$} for \mbox{0.03~$\mu$m $\le s \le 1$~mm}. We further assumed that the grains are spherical and homogeneous, composed of astronomical silicate \citep{Weingartner01a}.

Using the 3~Myr isochrone from \cite{Siess00a} we calculated the stars radii and temperatures from the sink masses. Importantly, as in \citetalias{Borchert22b}, we used the time-averaged mass accretion rate on the sink particles between two calls to \textsc{mcfost} to calculate the accretion luminosity at the stellar surface (2.0~R$_{\odot}$ for the primary and 1.3~R$_{\odot}$ for the secondary). This accretion luminosity was added to the stellar luminosity assuming the accretion luminosity is released over the whole stellar surface and emits as a black-body. Although FU Ori is likely younger than 3~Myr, this assumption has no strong effect. A younger age would lead to cooler effective temperatures and larger stars, but assuming 2~Myr would only change the accretion luminosity by $\sim$~10\% (with radii of 2.3~R$_{\odot}$ for the primary and 1.5~R$_{\odot}$ for the secondary). We performed an additional simulation using 2~Myr as the stellar age and confirmed that the effect on our overall results is negligible.

For our Monte Carlo radiative transfer we set the number of photons to $1.28\times10^8$. After \textsc{mcfost} has finished the temperature calculations, we assume that the gas temperature is equal to the dust temperature and use this as the gas temperature in \textsc{phantom} for subsequent evolution. We did not include PdV work or shock heating in our initial set of simulations. To test this assumption, we performed a comparison including the PdV work and shock heating as a source term for the radiative transfer in one of our \textsc{mcfost} coupled simulations. We found the contribution from PdV and shock heating negligible, as the accretion luminosity dominates the heating process in our simulations, see Section~\ref{sec:temperatures} and Appendix~\ref{app:pdv} for more detail.

We used mostly the same initial setup presented in \citetalias{Borchert22b}. The periastron distance was set to 20 au and the initial separation between both stars was at 10 times periastron, i.e. \mbox{200 au}. Both stars were treated as sink particles with an accretion radius of 0.5 au, meaning that SPH particles located within this radius are accreted by the sink particle (conditionally within 0.5 au and unconditionally within \mbox{0.4 au}; see \citealt{Price18b}). We set the primary mass to $1.4~{\rm M_{\odot}}$ with a surrounding disc of $0.02~{\rm M_{\odot}}$, an inner disc radius at 1 au and an outer disc radius at \mbox{50 au}. We assumed the secondary had a mass of $0.5~{\rm M_{\odot}}$  with a surrounding disc of $0.007~{\rm M_{\odot}}$, an inner disc radius at 1 au and an outer disc radius at 17 au. The initial surface density profile for each disc was \mbox{$\Sigma(r) \approx r^{-1}$} with an initial $H/R=0.05(R/1~{\rm au})^{1/4}$ for particle placing that was subject to change following temperature calculations during radiative transfer. To mimic a disc viscosity we used the SPH shock viscosity \mbox{$\alpha_{\rm av}=0.1$}, giving $\alpha_{\rm disc}\approx 0.0023$ \citep{Lodato10a}, typical of non-outbursting protoplanetary discs. $10^6$ equal mass SPH particles were used in our simulations, initially spread over both discs.

With two discs in the simulation that can rotate in two directions, we have four possible combinations of disc rotations. The perturber can be on an inclined prograde ($\beta=45\degree$) or inclined retrograde ($\beta=225\degree$) trajectory. Additionally, the perturber disc can either rotate in the same or opposite relative direction as the primary disc.

We computed the disc masses and radii of both stars at the start and end of the simulation, averaging over the first and last \mbox{15 years} of the simulation runs. The mass was determined to include all particles with a density \mbox{$>10^{-13}$ g cm$^{-3}$} within a radius chosen to be 1.2 times the initial set radius of the disc in order to include all particles from the viscous expansion. Radii were defined to be the average radius containing 99\% of the disc mass. Disc masses and radii are listed in Table~\ref{tab:mass}.

\vspace{-0.25cm}
\section{Results}\label{sec:results}
\begin{figure*}
    \centering
    \includegraphics[width=\textwidth]{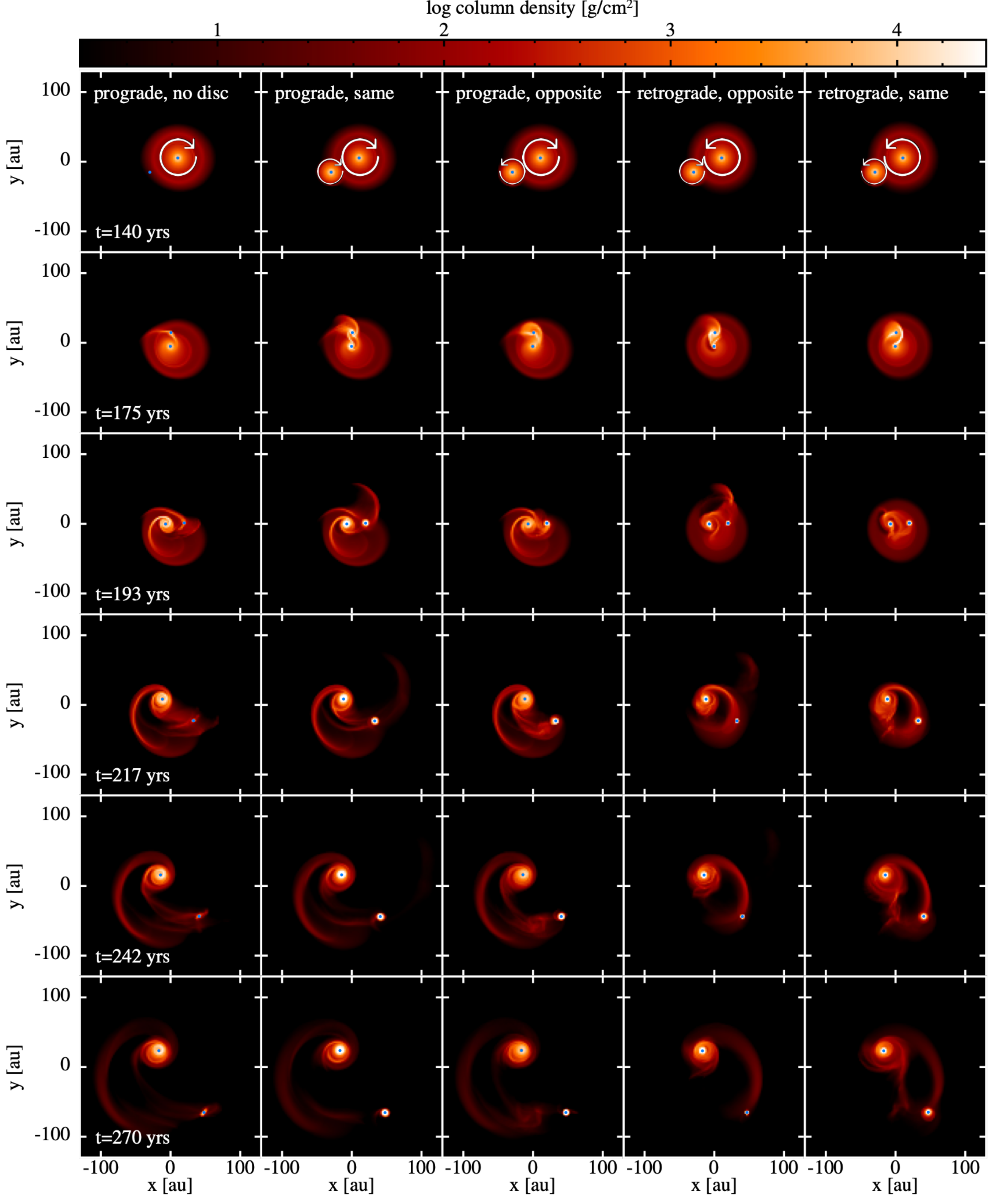}
    \caption{Column density evolution of disc-disc flybys, viewed face-on to the initial disc rotation. Each row shows a different time in the simulation: before (140 years), just before (175 years) and after (193+ years) periastron. Each column shows a different simulation with the differentiating features listed in the first panel of each column. Arrows indicate the rotations of the disc. Blue dots indicate the sink particles. Both discs show disc truncation by tidal interaction. The secondary disc ends up more compact in simulations where initial disc rotations are opposite.}
    \label{fig:density}
\end{figure*}

\begin{table*}
    \centering
    \begin{tabular}{l p{1cm} p{1.5cm} p{1cm} p{1.3cm} p{1cm} p{1.5cm} p{1cm} p{1.3cm}}
    \hline
    Simulation & 
    $R_{1.f}$ [au] & 
    $M_{1,{\rm disc},{\rm f}}$ [$10^{-3}$~M$_{\odot}$] & 
    $M_{1,{\rm acc}}$ [$10^{-3}$~M$_{\odot}$] & 
    \%$M_{1,{\rm acc}}$ from disc 2 &
    $R_{2.f}$ [au] &
    $M_{2,{\rm disc},{\rm f}}$ [$10^{-3}$~M$_{\odot}$] &
    $M_{2,{\rm acc}}$ [$10^{-3}$~M$_{\odot}$] & 
    \%$M_{2,{\rm acc}}$ from disc 1\\
    \hline
    \hline
    prograde, no disc & 24.6 & 8 & 0.2 & -- &  
    10.1 & 1.4 & 1.5 & 100\% \\
    prograde, same & 24.0 & 10 & 0.5 & 55\% & 
    6.0 & 3.9 & 2.6 & 63\% \\
    prograde, opposite & 24.7 & 9 & 0.3 & 9\% &
    4.0 & 3.6 & 5.3 & 47\% \\
    retrograde, opposite & 21.1 & 12 & 2.3 & 54\% &
    3.9 & 0.6 & 3.6 & 26\%\\
    retrograde, same & 25.4 & 12 & 2.3 & 33\% &
    10.9 & 2.8 & 2.0 & 21\% \\
    \hline
    \end{tabular}
    \caption{Disc radii and masses for the primary and secondary discs at the end ($f$) of the simulations as well as the accreted mass onto each star and the fraction of material accreted from the disc initially around the \emph{other} star. Initial disc radii and masses were 32.9~au and $12 \times10^{-3}$~M$_{\odot}$ for the primary and 16.8~au and $6.9 \times10^{-3}$~M$_{\odot}$ for the secondary. Disc masses are defined as all particles with a density \mbox{$>10^{-13}$~g~cm$^{-3}$} within a radius chosen to be 1.2 times the initial radius of the disc. Radii are defined to be the average radius containing 99\% of the disc mass. All values are the averages over the first and last 15 years of the simulations.}
    \label{tab:mass}
\end{table*}

\begin{figure*}
    \centering
    \includegraphics[width=\textwidth]{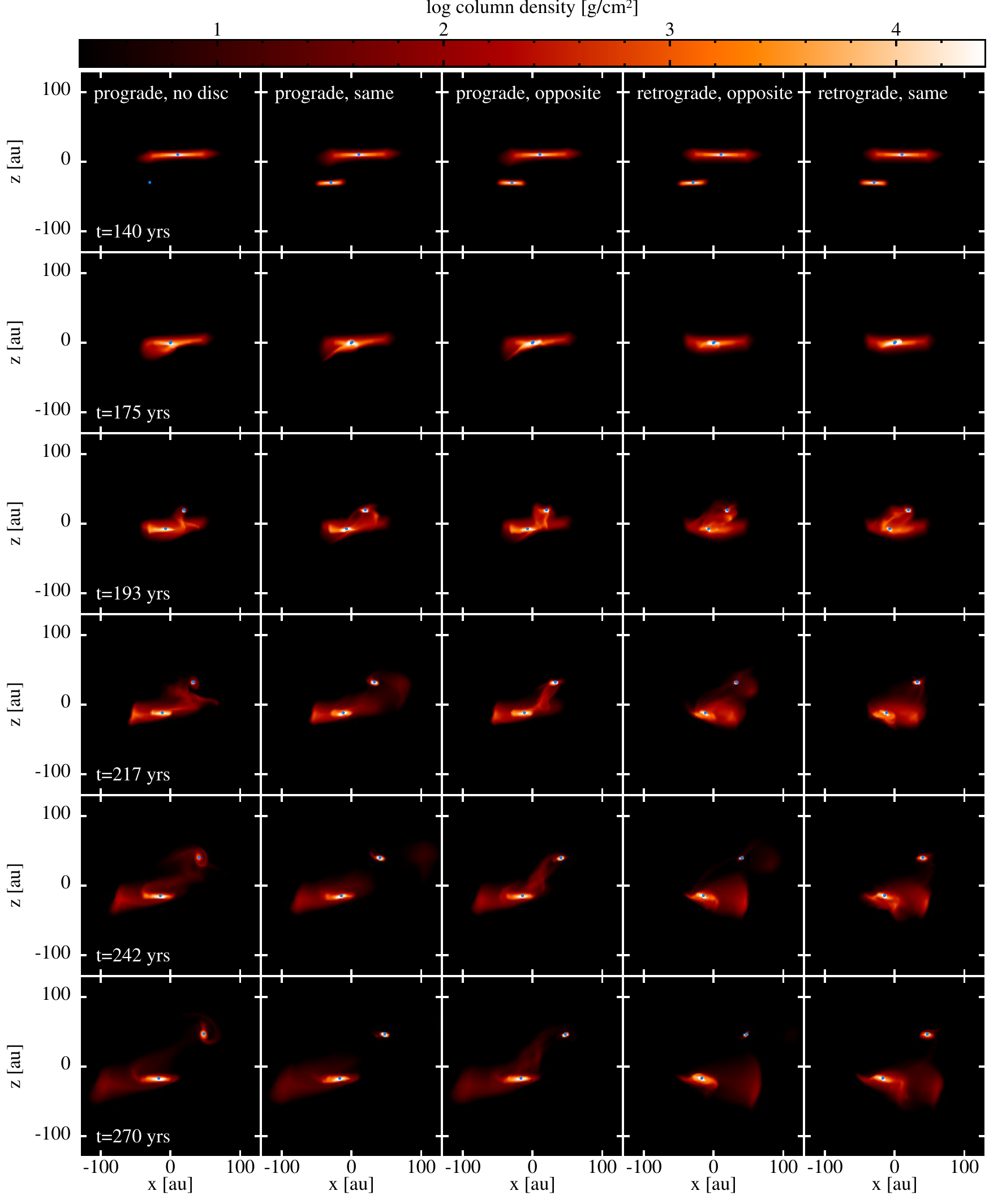}
    \caption{Side view column density evolution. The panels represent the same simulations as in Figure~\ref{fig:density}. Both discs are warped following the encounter, with retrograde encounters leading to larger change in inclination of the primary disc compared to prograde encounters. This could explain how the orbital plane of planetary systems can be misaligned with respect to the stellar rotation axis.}
    \label{fig:density-side}
\end{figure*}

\begin{figure*}
    \centering
    \includegraphics[width=1\textwidth]{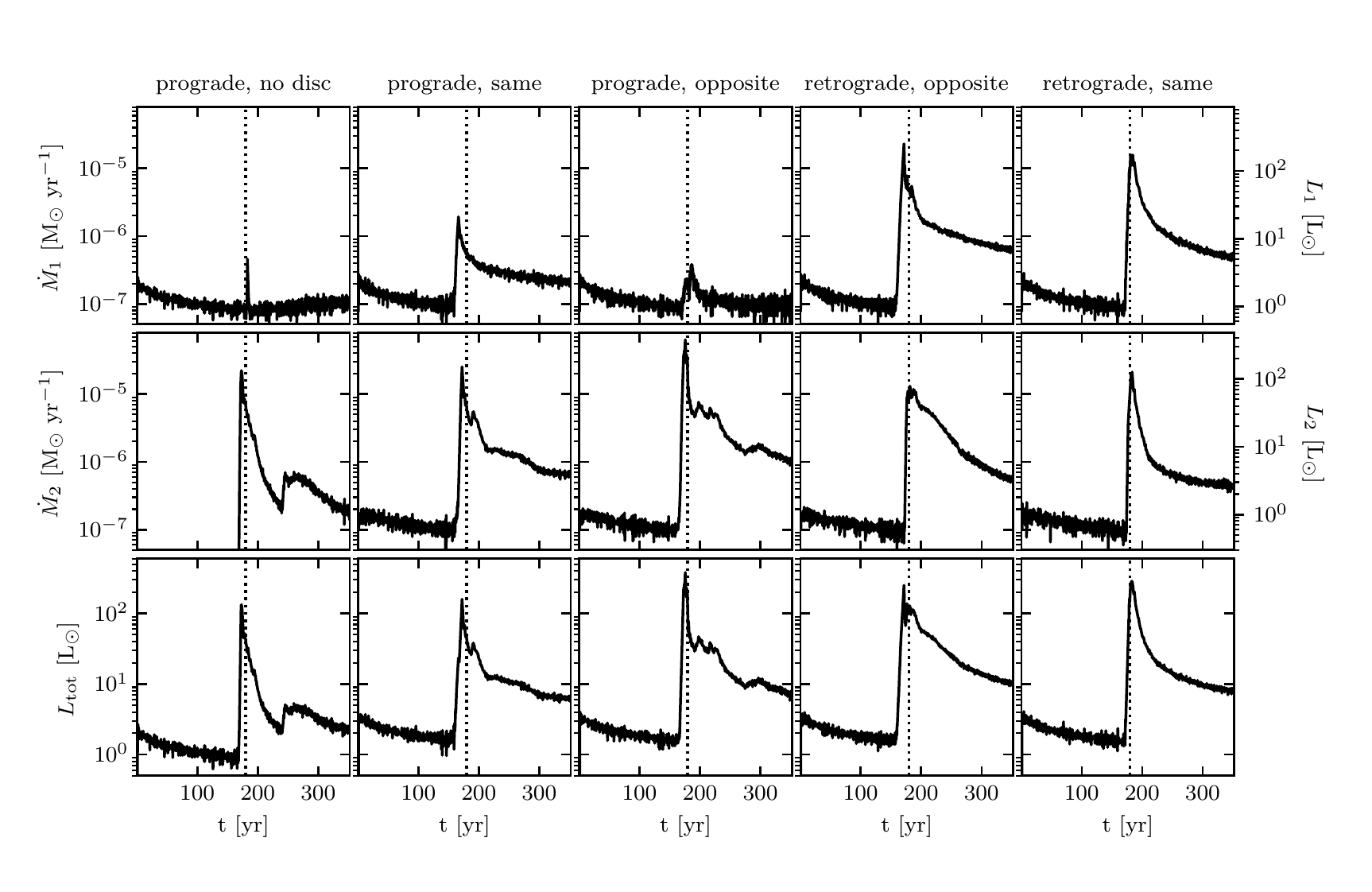}
    \caption{Accretion rates as a function of time for the different disc-disc flybys showing the primary (top row) and secondary (middle row) mass accretion rate with the appropriate accretion luminosity on the right axis. The bottom row shows the total accretion luminosity of both stars added together. The vertical dotted line indicates when periastron occurred at 179.5 years. Overall, the FU Orionis-type outburst, either in the secondary (prograde encounter) or in both the primary and secondary (retrograde encounter), is 2--10 times longer when a pre-existing circumstellar disc is present (comparing left column to right four columns).}
    \label{fig:mdot}
\end{figure*}

\begin{figure*}
    \centering
    \includegraphics[width=1\textwidth]{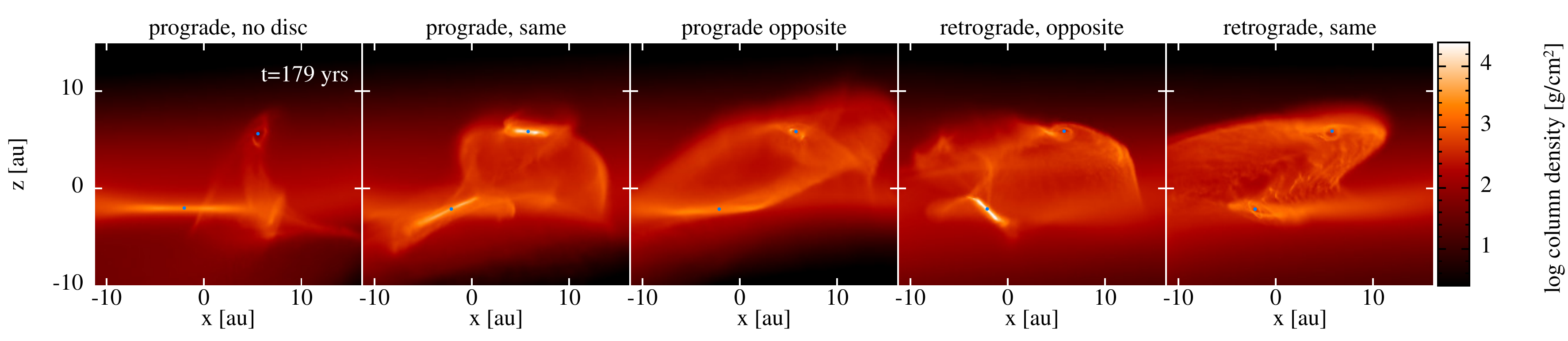}
    \caption{Side view column density zoomed in on the interaction between both stars. The panels each show a snapshot from a different simulation. The primary is in the lower left and the secondary in the upper right corner of each panel. There is misaligned material present which causes the star(s) to go into outburst.}
    \label{fig:zoom}
\end{figure*}

\begin{figure*}
    \centering
    \includegraphics[width=\textwidth]{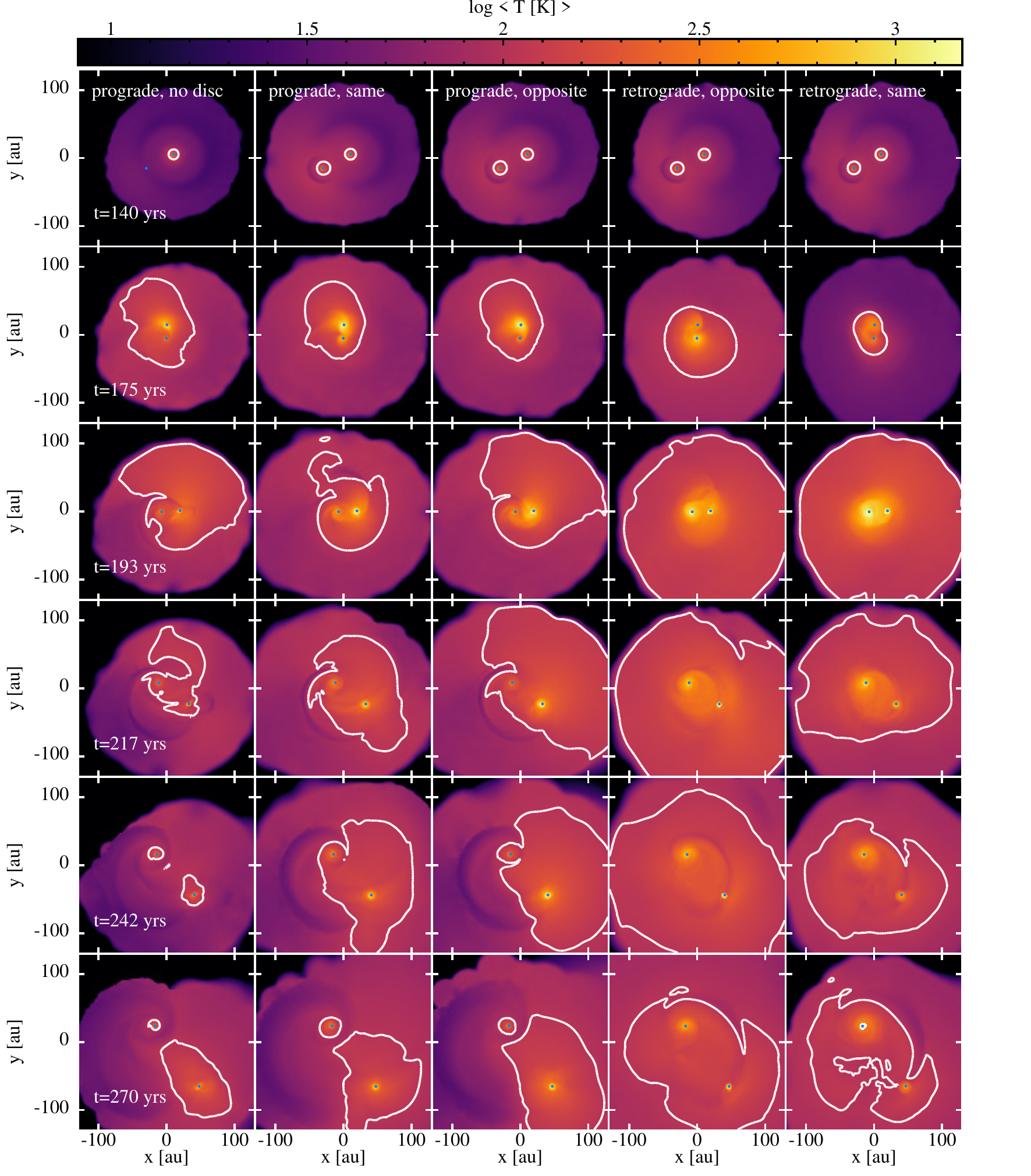}
    \caption{As in Figure~\ref{fig:density}, but showing temperature evolution with panels showing density weighted averaged temperatures along the line of sight. Temperatures exceed 1500 K during the encounter if a pre-existing circumstellar disc is present, with the outburst occurring primarily in the secondary disc (prograde encounter) or the primary disc (retrograde encounter). The contour shows the location of the iso-temperature line corresponding to the water snowline with a temperature of $105$~K \citep{Cieza16a}.}
    \label{fig:temperature}
\end{figure*}

\subsection{Disc evolution}
Figure~\ref{fig:density} shows the evolution of column density in the x-y plane corresponding to the original disc plane of one simulation without a circumsecondary disc and 4 different disc-disc flybys. Each column represents a different simulation, with the defining differences titled above each column. Arrows in the first row indicate the rotations of the discs. The rows show different times in the simulations: Before (140 years), just before (175 years) and after (193+ years) periastron, which occurs at 179.5 years. The time is since the start of the simulation. Overall, both discs are truncated by the tidal interaction, though where disc rotations are in opposite directions, the perturber disc is more compact post the encounter than in simulations where both discs rotate in the same direction. This is reflected in the final disc sizes in Table~\ref{tab:mass}. The primary disc develops one to two large scale spiral arms following the encounter.

Figure~\ref{fig:density-side} shows a side view of the same simulations presented in Figure~\ref{fig:density}. Post periastron, the main disc is slightly inclined for prograde flybys by $\approx -10\degree$, and for retrograde flybys by $\approx +20\degree$ compared to the initial disc inclination (corresponding to the x-y plane in our simulations). 

\vspace{-0.25cm}
\subsection{Mass accretion rate}
Figure~\ref{fig:mdot} shows the mass accretion rates corresponding to the simulations displayed in Figures~\ref{fig:density} \& \ref{fig:density-side} with each column displaying a different simulation. The mass accretion rates are split up into primary (top row) and secondary (middle row) mass accretion rates.  In the case without a pre-existing circumsecondary disc, the secondary has no mass accretion rate prior to the encounter, as there is nothing to accrete. The right axes of the mass accretion rate rows show the corresponding accretion luminosity of the star. The bottom row shows the total accretion luminosity of both stars added together.

In prograde scenarios the secondary experiences the main outburst, while both stars experience similar sized outbursts in retrograde scenarios  of disc-disc encounters. In a retrograde case with no perturber disc, while the outburst on the primary is higher than for prograde cases, it is less than an order of magnitude in size and also an order of magnitude shorter than the outburst on the perturber. In all simulations, the star experiencing the main outburst reaches a maximum amplitude in the mass accretion rate of about $2$--$5.5\times10^{-5}~{\rm M_{\odot}~ yr^{-1}}$, 200--550 times higher than their initial rate. Within the first 50 years past the outburst, the mass accretion rate drops by a factor $\sim$10, especially for prograde encounters. Throughout the length of the simulations, the simulations including a pre-existing circumsecondary disc sustain the high accretion rate for 2--5 times longer. The mass accretion rate of the secondary in the no-secondary-disc case drops by a factor of 10 in the first 15 years, and a further factor of 8 in the next 35 years. The secondary's mass accretion rate drops even faster for a retrograde flyby with no circumstellar disc present. The other simulations showing mass accretion rates which are still $>10$ times higher and continuing to end up 2--4 times higher at the end of the simulations compared to the no disc scenario. These higher rates stay between $(1-5)\times 10^{-6}~{\rm M_{\odot}~ yr^{-1}}$ for 50--100 years after the outburst before finally dropping to $\ge 8\times 10^{-7}~{\rm M_{\odot}~ yr^{-1}}$ at the end of the simulation, $\sim$180~yrs past periastron. The high mass accretion rate is driven by misaligned flows with material raining down onto the star from different directions (see Figure~\ref{fig:zoom}). Once material settles back into a single plane with a well-defined angular momentum direction, the accretion rate returns to pre-burst levels.

In Table~\ref{tab:mass} we quantify how much of the mass accreted onto a star is accreted from material initially in its own disc, or from the other star's disc over the length of the simulation. In the cases where the star experiences an outburst, a large fraction of the accreted mass ($\gtrsim 20$\%) consists of alien material that initially resided in the disc around the other star. The primary \emph{only} goes into outburst when there is significant material captured from the circumsecondary disc onto the primary, resulting in cancellation of angular momentum and rapid accretion (as seen in Figure~\ref{fig:zoom}).

\vspace{-0.25cm}
\subsection{Disc temperatures}
\label{sec:temperatures}
Figure~\ref{fig:temperature} shows a face-on view of the temperature evolution of the same simulations presented in the previous figures. The second row shows that the disc temperature rises to high temperatures (above 1000 K) just before periastron for prograde flybys, while it is delayed for retrograde flybys. In the retrograde case, the primary disc stays heated for a longer period of time post periastron while it cools down quickly for prograde flybys. For the secondary, the surrounding disc stays more heated for flyby scenarios where the disc rotations are opposite to that of the primary disc. 

Temperatures in the discs rise to above 1500 K in simulations with a pre-existing circumsecondary disc. In the prograde scenarios, the secondary experiences temperatures of more than 1500 K in an circle with a $\sim$2 au radius around it --- which persists for decades --- while the primary experiences temperatures of up to 1500~K within 1~au for less than 3 years. In retrograde cases, both stars have discs with more than 1500 K surrounding them, with the disc area experiencing such high temperatures around the primary having a radius of \mbox{$\sim$6 au}, and around the secondary of \mbox{$\sim$3 au}. In contrast, the simulation of the star-disc encounter sees temperatures only reaching \mbox{$\sim$ 1000 K} in a radius of 1.5 au around the secondary as it passes through the circumprimary disc. Temperatures this high should in either case sublimate dust present in the discs, with affected areas varying.

In our simulations the main source for the temperature is passive heating from accretion luminosity radiated from the stellar surfaces. Active disc heating is negligible because we are not modelling the inner regions of the disc at $< 0.5$~au. Our inner discs start at $\sim$~25--40 times the stars' radius. \cite{Zhu07a} showed that most of the luminosity in FU Orionis is coming from accretion luminosity at the stellar surface (see $R=1R_i$ curve in figure~4 of \citealt{Zhu07a}) but with significant contribution from the inner disc (1--5 times the stellar radius), which means that shock heating would become more important when the inner disc is modelled. We show a comparison simulation including PdV work and shock heating for the prograde case where both discs rotate in the same direction (see Figure~\ref{fig:pdv} in Appendix~\ref{app:pdv}).

The contour in Figure~\ref{fig:temperature} shows the approximate location of the iso-temperature line corresponding to the water snow-line (region within a protoplanetary disc where the condensation temperature of a major volatile is reached) with a temperature of 105~K \citep{Cieza16a}. The iso-temperature line can be seen to shift from an initial $\sim$10 au pre-encounter to up to $\sim$100 au during the encounter.

\vspace{-0.25cm}
\section{Discussion}\label{sec:discussion}
\begin{figure*}
    \centering
    \includegraphics[width=\textwidth]{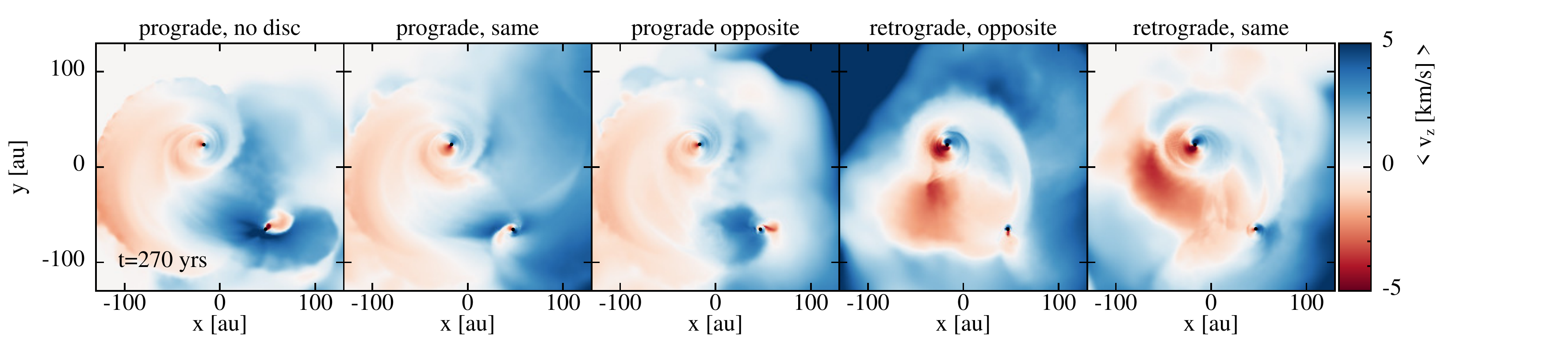}
    \caption{Column averaged line of sight velocity maps for the simulations shown in Figure~\ref{fig:density} $\sim 90$ years post periastron with $v_{\rm z}=0$~km/s as the velocity of the primary (located towards the upper left). The kinematics are a mess. We find expanding `bipolar shells' of material solely due to the interactions of both stars and discs during the encounter. Prograde and retrograde encounters are best distinguished in kinematics by the amplitudes and orientations in the velocity fields. Individual channel maps are shown for the middle simulation in Figure~\ref{fig:map}.}
    \label{fig:vel}
\end{figure*}
\begin{figure*}
    \centering
    \includegraphics[width=\textwidth]{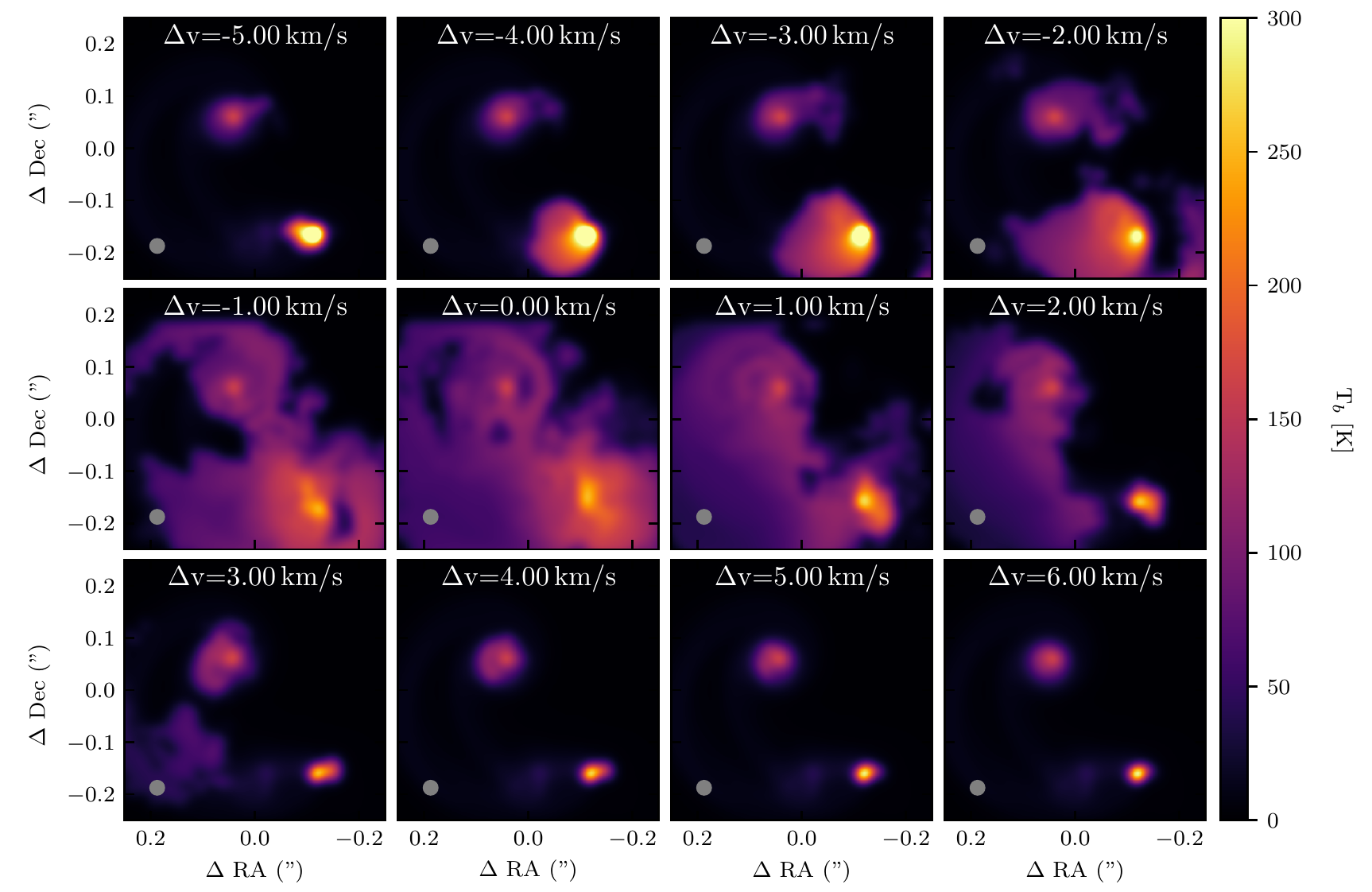}
    \caption{Synthetic CO(1-0) emission channel maps for the prograde flyby with opposite disc rotations at $t=270$~years. $\Delta {\rm v}$ indicates the change in the line of sight velocity from the systemic velocity set to the velocity of the primary (located in the upper left of the panel). No coherent disc rotations are obvious in the channel maps because of the disturbed kinematics. Typical line-of-sight motions of order several km/s occur, which could easily be mistaken for outflows.}
    \label{fig:map}
\end{figure*}
\subsection{Sustained outbursts}
Our simulations indicate that a pre-existing circumsecondary disc leads to a more sustained outburst compared to the case where no disc is present around the perturber. In \citetalias{Borchert22b} we showed that a stellar flyby creates a fast rise in the mass accretion rate, defined as the time it takes from pre-outburst accretion levels to the maximum $\dot{M}$, similar to what is seen in FU Orionis objects. One caveat was that the outburst wasn't maintained as long and high as observed in FU Ori itself, where the luminosity has remained approximately constant for 85+ years. By contrast, in \citetalias{Borchert22b} the mass accretion rate dropped $\sim$1.5 orders of magnitude within $<50$ years after the outburst.

Figure~\ref{fig:mdot} shows that stellar flyby simulations with pre-existing discs lead to a sustained outburst that is 2-10 times higher than without a circumsecondary disc pre-encounter, either in the primary or the secondary. Without a circumsecondary disc, the accretion rate of the secondary drops nearly two orders of magnitude after $\sim 50$ years, then experiences a slight jump up again before decreasing while the primary's accretion rate drops immediately after the outburst back to the previous level. 

The scenarios with a pre-existing circumsecondary disc show a drop of nearly a factor 10 in $\dot{M}$ for $\sim50$ years after the initial outburst. The mass accretion rate of either the primary or secondary then continues at a higher rate for the rest of the simulations (\mbox{100+ years}). After the first 50 years post outburst, $\dot{M}$ stays between \mbox{$(1-5)\times 10^{-6}~{\rm M_{\odot}~ yr^{-1}}$} for the next $\sim50$ years before dropping down to $\ge 8\times 10^{-7}~{\rm M_{\odot}~ yr^{-1}}$. 

We find two possible scenarios from a disc-disc encounter: both stars or only the perturber going into outburst. Which one occurs depends on the trajectory of the flyby (retrograde or prograde). In prograde encounters only the secondary goes into outburst while in retrograde encounters outbursts occur around both stars. In retrograde cases sometimes the primary very slightly dominates the outburst. The primary star accreting more mass during retrograde encounters aligns with findings in \cite{Kuffmeier21a}, who simulated a cloudlet of gas encountering a star-disc system and found the star accreting more mass in retrograde encounters due to the transport of angular momentum. One of the main results from our previous work in \citetalias{Borchert22b} was the discovery that the secondary dominates the outburst in a disc-penetrating stellar flyby. Most FU Orionis objects have not yet been discovered to have a companion which could have been a past flyby, and it is therefore not known if they will all show the perturber in outburst. 

Figure~\ref{fig:zoom} shows that there is a substantial amount of material around both stars that is misaligned with respect to the original plane of the corresponding disc(s). Outbursts are driven and sustained by misaligned material (mostly from the other star; see Table~\ref{tab:mass}) which leads to rapid change in the total angular momentum of the disc. The outburst continues so long as the star continues to have misaligned material `feeding' onto their disc. The substantial amount of alien material that is accreted during the encounter may lead to observable chemical abundance anomalies (such as the $\lambda$ Boo phenomena; \citealt{Murphy17a}).

\subsection{Observational implications}
Overall, we see that if we have a pre-existing circumsecondary disc in a disc penetrating stellar flyby, the mass accretion rate outburst, either on the primary or secondary, is maintained 2-10 times longer $\sim50+$~yrs past periastron compared to the scenario without a pre-existing circumsecondary. Translating our mass accretion rates into lightcurves assumes that nothing is obscured. It is probable that parts or all of the star and disc are obscured via extinction through dust, meaning that the observed lightcurves would differ. 

Looking at other available sources which are classified as FUors or EXors (table 1 in \cite{Audard14a}), we notice that FU Ori is an outlier in the group with a continuing outburst of nearly 100 years. The majority of objects listed have outburst durations of less than \mbox{30 years}, in some cases due to recent observations with the outburst still ongoing, in other cases with the onset of the outburst being further ago than the duration of the outburst listed. One example of this is V1057 Cyg, where the outburst had a duration of 10 years which started in 1970 and the early decline of this outburst was already visible in the lightcurve in figure 3 in \cite{Hartmann96a}. From our simulation results and the behaviour of other FUor/EXors it might even be the anticipated behaviour that the outburst declines within a few decades and could even be a future event to be observed in FU Ori itself.

From lightcurves alone it will be difficult to determine what kind of scenario played out in an FU Orionis event. So how can we distinguish between the cases? While one can make a guess based on which star is in outburst, a better approach is to look at the line of sight velocity maps. Figure~\ref{fig:vel} shows different simulations display different kinematic structures which can be compared to the kinematics seen in actual observations \cite[e.g.][]{Perez20a}. A simple prediction from our model is that kinematics in outbursting systems should be messy. Contrary to the suggestions in \cite{Ruiz-Rodriguez17a, Ruiz-Rodriguez17b} and \cite{Zurlo17a} that such velocity maps indicate conical outflows, we do not launch any outflows in our simulations and everything seen in Figure~\ref{fig:vel} is purely from the stellar flyby and the star-disc and disc-disc interactions. This was previously shown in flyby simulations by \cite{Cuello20a}. They observed arc-like features which are non coplanar to the rest of the disc in their channel maps (their figure 4) that appear as parts of a conical outflow, with no outflow involved. This does not imply that other outflows are not present, merely that they are not required to explain the observed few km/s flows.

Figure~\ref{fig:map} shows the prediction from our model in individual channels, showing synthetic CO(1-0) emission of the prograde simulation with disc rotations in the opposite direction. The primary is located in the top left of the panel, while the perturber is in the bottom right area. Again, the main story is that flyby kinematics are messy. We see line-of-sight motions of order $\Delta v=-5$--6~km/s, matching with our line of sight velocity map in Figure~\ref{fig:vel}. No coherent disc rotations are obvious on these scales.

\vspace{-0.25cm}
\subsection{High temperature processing}

The recorded high temperatures are interesting. Temperatures exceeding 1000 K will result in the dust sublimating. This sublimation could possibly be partly responsible for the creation of chondrules (requiring $\ge 1000$ K) or Calcium-Aluminium-rich inclusions (CAIs) (requiring $\ge 1500$ K) in our solar system, which formed \mbox{$\sim4.567$ Gyrs} ago \citep{Connelly12a}. Common theories for the formation of chondrules and CAIs rely on circulation or shock heating \cite{Scott07a} which has been suggested by theoretical analysis to be too variable for reliable formation \citep{Liffman09a,Stammler14a}. In the case of a stellar flyby, the CAIs should form naturally all over the disc at the same time as opposed to centrifugal ejection and re-entry in the disc \citep{Liffman16a}, where for a limited period of time you eject and form CAIs in the Solar Nebula.
One caveat to this in our simulations is that dust sublimation is not yet treated correctly in our models, though preliminary tests of removing the dust for the affected particles showed no changes in the disc temperatures. Furthermore, the radiative transfer calculations assume radiative equilibrium, which may not be the case. 

Additionally, we predict that the iso-temperature line corresponding to the water snow-line ($T_\mathrm{dust}=105$~K \citep{Cieza16a}) should move during the outburst (Figure~\ref{fig:temperature}). Interestingly, \cite{Cieza16a} inferred from observations of V883 Ori, an FU Orionis object, that the water snow-line moves from $\sim$5 au for non-outbursting solar-type stars to $>40$ au during protostellar accretion outbursts, more than 10 times the radius expected in passively heated discs \citep{Ruiz-Rodriguez17a}. While we did not include ices in our calculations, the motion of the $T=105$~K iso-temperature line in our simulations is consistent with these observations. The water snow-line has a non-circular shape, which has also been found for outbursts in gravitationally unstable discs \cite{Vorobyov21b}. This is a crude comparison though as in practice the snow-line depends on the temperature as well as the pressure \citep[see e.g. Figure 3 in][]{Okuzumi16a}, a complication beyond the scope of the present discussion.

High temperatures that occur during encounters between circumstellar discs might explain the unusual and complex chemistry observed in IRAS16293-2422 \citep[][and more]{Drozdovskaya18a, Murillo18a,van-der-Wiel19a}, which consists of a pair of interacting discs.

\subsection{Disc inclination}
The orientation of the secondary disc post-flyby in the disc-disc scenarios appears to be random. The final disc tilt changed by \mbox{10--45{\degree}} compared to the initial orientation. Such changes in the disc inclination can explain why the orbital plane of planetary systems are often misaligned with respect to the stellar rotation axis, such as the $\approx 7\degree$ obliquity in our solar system. The severity of the inclination of the captured and disturbed disc depends on the inclination of the flyby though the periastron distance or disc size would also have an effect on this.

A caveat is that our simulations were only evolved for a few hundred years post-periastron. To confirm that the disc inclinations persist, we thus evolved the simulations to $t\approx 2000$~years and observed that both discs remain tilted. The tilting in the primary decreases by $\sim$~2--10~\% while the secondary disc experiences changes of +97 to -40~\% in the tilt over this timescale.

\begin{figure}
    \centering
    \includegraphics[width=\columnwidth]{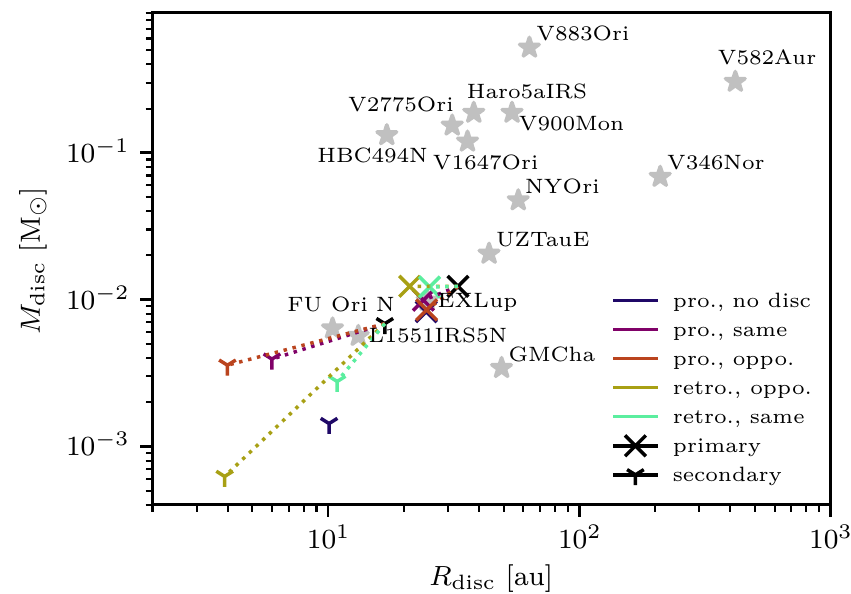}
    \caption{Disc mass vs. disc radius at the end of our simulations (from Table~\ref{tab:mass}). Different colours represent different simulations with a black marker indicating the initial disc mass and radius while different markers distinguish between primary and secondary of the simulation. Lines have been added to show the move from the initial to final disc masses and radii. The grey stars represent disc masses and radii for FUor/EXor sources \citep[taken from figure 6 in][]{Hales20a}. We assumed FU Ori-like initial disc masses but it should be noted that disc masses of most FUors are much higher than this.}
    \label{fig:massradius}
\end{figure}

\vspace{-0.25cm}
\subsection{Disc truncation}
\cite{Hales20a} suggested that disc masses and radii for various FUor/Exor may differ from typical Class I and II disc populations. Figure~\ref{fig:massradius} compares the observed final disc masses and radii from our simulations to measurements of FUor/Exor objects from \cite{Hales20a}. Our simulations assume FU Ori-like disc masses, but observed disc masses are much higher for most FUors \citep{Kospal21a}. Higher disc masses would lead to correspondingly higher mass accretion rates.
The main effect of the flyby is to truncate both discs, as seen by the dotted lines connecting our initial and final disc sizes and masses. 

\vspace{-0.25cm}
\subsection{Likelihood of close encounter}
While a close stellar flyby can explain the sudden FU Orionis outburst, we need to also look into the likelihood of such a close encounter. \cite{Clarke91a} derived the collision rate through a disc at distance $R_{\rm disc}$ as

\begin{equation}
    \Gamma_{\rm hit} = \Gamma_0 \left( 1 + \frac{V_*^2 R_{\rm disc}}{G M_*}\right) \ \ \ \mathrm{with}\ \ \  
    \Gamma_0 = \frac{4 \sqrt{\pi}n_0 G M_* R_{\rm disc}}{V_*}.
\end{equation}

Representing a dense stellar environment we use $n_0=100$~pc$^{-3}$ and $V_*=$1~km/s \citep{Binney87a,Clarke91a}. With an encounter distance of $R_{\rm disc}=20$~au and a perturber mass of $M_*=0.5~{\rm M_{\odot}}$ we derive a collision rate of $\Gamma_{\rm hit} = 1.57\times 10^{-4}$~Myr$^{-1}$. The probability for an individual star to have experienced such an encounter in the past 100 years is therefore \mbox{$P_{\rm hit}=\Gamma_{\rm hit} \times 10^{-4}~{\rm Myr} = 1.57\times 10^{-8}$}. Using this encounter probability, the probability of a flyby occurring in the past 100 years in a cluster of a given size is
\begin{equation}
    P_{\rm flyby} = 1 - (1-P_{\rm hit})^{n_{\rm pairs}},
\end{equation}
with $n_{\rm pairs}=n_{\rm stars} \times (n_{\rm stars}-1)/2$ where $n_{\rm stars}$ is the number of stars in the cluster and $n_{\rm pairs}$ the number of possible pairs.

We thus arrive at a probability of $\approx9$\% for a 20 au encounter in the last 100 years in a young cluster of $\approx$3500 stars \citep[e.g. the Orion Nebula Cluster,][]{Hillenbrand98a}. This simplified calculation does not take into account the possible range of disc radii, perturber masses, cluster sizes or stellar density gradients, though it provides an estimate of the likelihood for such events \citep[see also][]{Cuello22a}. Regardless of whether or not such rate estimates are accurate, at least two of the known FU Ori objects (FU Ori; \citealt{Perez20a} and Z CMa; \citealt{Dong22a}) show strong evidence for recent stellar flybys.

\vspace{-0.25cm}
\section{Conclusions}\label{sec:conclusion}
We performed a set of experiments of disc-penetrating stellar flybys in young stars in the case where there are pre-existing circumstellar discs around both stars prior to the encounter. We explored prograde and retrograde disc rotations with respect to the binary orbit. Our conclusions are as follows:

\begin{enumerate}
    \item FU Orionis-type outbursts are sustained for longer and with higher peak $\dot{M}$ when a pre-existing disc is present around the perturber. 10 times higher in the first \mbox{60 years} after the encounter ($\sim 5-10\times10^{-6}~{\rm M_{\odot}~yr^{-1}}$) and then dropping to 2-4 times the no disc outburst at the end of the simulation ($\sim 8-10\times10^{-7}~{\rm M_{\odot}~ yr^{-1}}$).
    \item Which star experiences the main outburst depends on the flyby trajectory. In prograde flybys, where the orbit is in the same direction to the circumprimary disc rotation, the secondary experiences the main outburst. In retrograde flybys, where the orbit is in the opposite direction to the rotation of the circumprimary disc, both the primary and the secondary go into outburst. The amplitude of the main outburst in either case is of the same order, $\sim 2-5.5\times10^{-5}~{\rm M_{\odot}~ yr^{-1}}$.
    \item A large fraction (20--100\%) of the material accreted by either star during an outburst is captured from the disc around the other star. This shows that outbursts are mainly driven by fresh inflows of misaligned material leading to direct cancellation of angular momentum. The strength and duration of the outburst depends on the efficiency by which such material is captured.
    \item Pre-existing circumstellar discs lead to a further increase in disc temperatures during the flyby encounter, with inner discs at \mbox{2--6 au} of the outbursting star reaching > 1500 K instead of \mbox{$\sim$1000 K} within 1.5 au for the star-disc encounter.  
\end{enumerate}

Something we have not yet explored is the effects of dust evolution in our flyby scenarios with live temperature feedback. \cite{Cuello19a} explored dust evolution in non-penetrating flybys, but in our case sublimation will occur due to the temperatures exceeding \mbox{1000 K}. Modelling the dust disc(s) in our simulations would allow for a direct explanation of the compact mm-emission seen around both stars discs in FU Ori \citep{Perez20a}.

\vspace{-0.25cm}
\section*{Data availability}
\textsc{phantom}\footnote{\url{https://github.com/danieljprice/phantom}} and \textsc{mcfost}\footnote{\url{https://github.com/cpinte/mcfost}} are publicly available
and the simulation setup and data underlying this article will be shared on request.

\vspace{-0.25cm}	
\section*{Acknowledgements}
EMAB acknowledges an Australian Government Research Training Program (RTP) and a Monash International Tuition Scholarship. We acknowledge Australian Research Council grants FT170100040 and DP180104235. NC acknowledges European Union H2020 funding via Marie Skłodowska-Curie grant 896319 and ANR (Agence Nationale de la Recherche) of France under contract number ANR-22-ERCS-0002-01. We used Ozstar at Swinburne University and Gadi via the National Computing Infrastructure. We used \textsc{splash} \citep{Price07a}, \textsc{Matplotlib} \citep{matplotlib}, \textsc{NumPy} \citep{numpy} and \textsc{CMasher} \citep{cmasher}.



\vspace{-0.45cm}
\bibliographystyle{mnras}
\bibliography{phd}


\vspace{-0.25cm}
\appendix
\vspace{-0.25cm}
\section{Heating from PdV work} \label{app:pdv}
\begin{figure}
    \centering
    \includegraphics[width=\columnwidth]{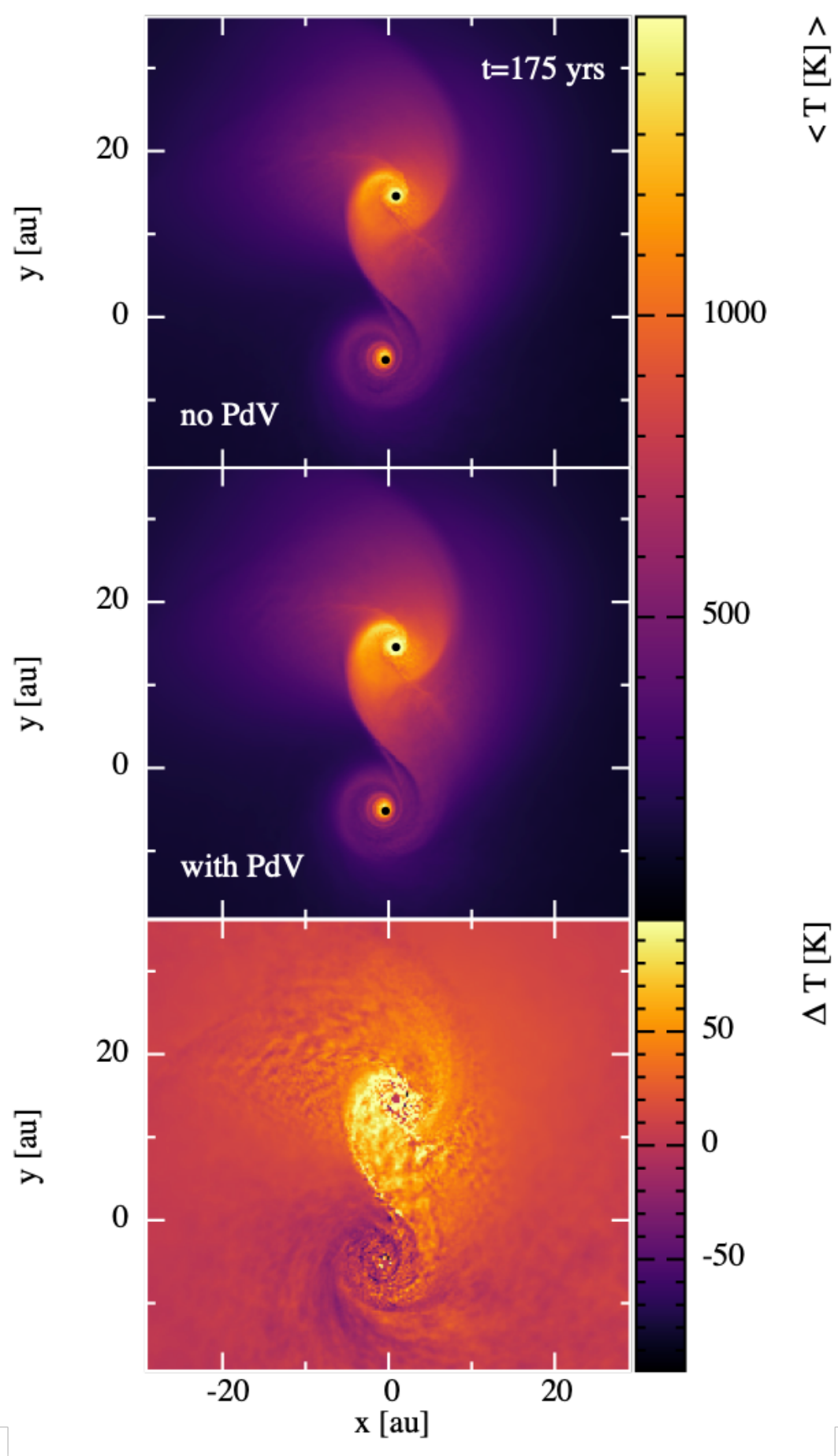}
    \caption{Prograde flyby with discs in the same orientation at periastron. The images show the difference between not including ({\it top}) and including ({\it middle}) the heating from PdV work and viscous/shock heating terms, and the difference between the two ({\it bottom}).}
    \label{fig:pdv}
\end{figure}
Figure~\ref{fig:pdv} compares two simulations of the prograde flyby without (top) and with (middle) the additional contributions from PdV work and viscous/shock heating included as source terms used by {\sc mcfost} when computing temperatures in the simulation. While the PdV work term makes up about $\sim 25$\% of the pre-outburst accretion luminosity, this amount is negligible during the encounter, making up less than 1~\% of the accretion luminosity. With shock heating included, shocks are up to 10\% warmer in the gas (bottom panel). However, the effects are small and do not change our overall results.

\bsp	
\label{lastpage}
\end{document}